\documentclass[journal]{IEEEtran}
\usepackage{amsmath,amsfonts}
\usepackage{algorithmic}
\usepackage{algorithm}
\usepackage{array}
\usepackage[caption=false,font=normalsize,labelfont=sf,textfont=sf]{subfig}
\usepackage{textcomp}
\usepackage{stfloats}
\usepackage{url}
\usepackage{verbatim}
\usepackage{graphicx}
\usepackage{cite}
\usepackage{cleveref}
\usepackage{bm}
\usepackage{amsmath}
\usepackage{empheq}
\usepackage{booktabs}
\usepackage[utf8]{inputenc}
\def\BibTeX{{\rm B\kern-.05em{\sc i\kern-.025em b}\kern-.08em
    T\kern-.1667em\lower.7ex\hbox{E}\kern-.125emX}}
\usepackage{balance}
\begin{document}

\title{Inertia Matching Principle: Improving Transient Synchronization Stability in Hybrid Power Systems With VSGs and SGs}

\author{
Changjun He,~\IEEEmembership{IEEE Member}, Li Zhang,~\IEEEmembership{Senior Member}, Qi Liu,~\IEEEmembership{IEEE Member}, Rui Zou
\thanks{This work was supported by the National Natural Science Foundation of China (U25B20203) and the Open Project of the Key Laboratory of Modern Power System Simulation Control $\And$ Green Power Technology, Ministry of Education (Northeast Electric Power University) (MPSS2026-0x). \textit{(Corresponding author: Li Zhang)}.}
\thanks{Changjun He, Li Zhang, and Qi Liu are with the School of Electrical and Power Engineering, Hohai University, Nanjing, China, 211100 (e-mail: hcj25@hhu.edu.cn; 20140074@hhu.edu.cn; lqhhu@hhu.edu.cn).}
\thanks{Rui Zou is with the State Grid Electric Power Research Institute, Nanjing, China, 211106 (e-mail: zourui@sgepri.sgcc.com.cn).}
}

\markboth{}%
{Shell \MakeLowercase{\textit{et al.}}: A Sample Article Using IEEEtran.cls for IEEE Journals}


\maketitle

\begin{abstract}
This paper investigates the transient synchronization stability in power systems hybridized with virtual synchronous generators (VSGs) and synchronous generators (SGs). A relative swing equation model is established to capture the transient synchronization dynamics between the VSG and the SG. Based on this model, both static and dynamic characteristics are systematically analyzed, and a quantitative stability level index is derived to elucidate the underlying stability mechanism. Then, two fundamental inertia matching principles are identified. First, a new instability mechanism induced by improper inertia matching between the VSG and the SG is revealed. It is identified that increasing the VSG's inertia does not monotonically improve transient stability, as commonly presumed. Instead, an optimal inertia matching constant exists that maximizes stability performance. Second, the influence of the VSG share on the synchronization stability is discovered to be strongly influenced by the matching between the VSG's inertia level and its voltage strength (i.e., output impedance). To achieve reliable and robust synchronization stability, proper coordination between the VSG's inertia and virtual impedance is essential. Finally, a coordinated stabilization strategy based on inertia matching and virtual impedance adjustment is proposed to enhance transient synchronization stability performance while suppressing fault current. Simulations conducted on a two-machine system and the IEEE 39-bus system validate the theoretical findings and demonstrate the effectiveness of the proposed strategy. 
\end{abstract}

\begin{IEEEkeywords}
hybrid system, virtual synchronous generator, inertia matching, transient synchronization stability.
\end{IEEEkeywords}

\section{Introduction}
\IEEEPARstart{A}{s} the penetration rate of renewable energy continues to increase, the grid faces severe challenges in maintaining transient synchronization stability. For example, a large-scale loss of synchronism (LOS) of grid-connected inverters occurred in the southeastern United States in 2023, causing nearly 1000 MW of photovoltaic generation to disconnect from the grid \cite{nerc20232023}.

Traditional grid-connected inverters employ grid-following (GFL) control strategies to maintain synchronization with the grid. However, due to the poor stability of the GFL in weak grids and its limited grid-support capability, grid-forming (GFM) control technologies have emerged in recent years as effective solutions \cite{li2022hierarchical}. Among these, virtual synchronous generator (VSG) control stands as one of the most mature approaches \cite{wang2025low}.

Lots of work has been done on the transient synchronization stability of the VSG coupled to the infinite grid. The allowable control parameter of the VSG to maintain transient stability is studied in \cite{chen2023quantitative}. How to quantitatively design the virtual inertia and damping is solved. While the VSG features voltage-source output characteristics, the fault current is relatively large during a voltage sag. Since the inverter has a weak current-tolerance capability, current-limiting strategies should be adopted to avoid damage. The strategies can be classified into two categories: current saturation limiter and virtual impedance control \cite{chen2023quantitative, liu2023dynamic}. Reference \cite{wang2023transient} analyzed the influence of current saturation on synchronization stability. It is identified that the current saturation loop reduces the stability region and worsens the stability. To suppress the current and improve the transient stability, reference \cite{li2022analysis} derives the virtual impedance boundaries and designs the optimal virtual impedance control strategy. Besides, regarding the multi-VSG scenario, the uniform and the nonuniform damping effects are identified using the phase portrait method in the paralleled VSG system \cite{qu2024transient}.

As the proportion of renewable energy increases, the power system will take on a form with inverter-based generation and synchronous generators (SG) co-existing. The transient synchronization stability of the hybrid system is co-dominated by the inverters' control algorithm and the rotors' dynamics, showing complicated mechanisms. Reference \cite{he2022transient, shen2020transient} investigates the transient stability in a hybrid system with GFL and SG. The inertia ratio between the GFL and the SG is found to play a vital role in the synchronization stability \cite{he2022transient}. Transient synchronization stability in the hybrid system with GFM and SG is analyzed in \cite{xue2024transient} by comprehensively considering the pre-fault, fault-on, and post-fault stages. However, the GFM is equipped with the droop control, which has little inertia and is therefore difficult to keep the frequency stability in low-inertia systems \cite{he2024analysis, feng2022provision}. Inertia, which is an important control parameter of the VSG and an inherent characteristic of SG, has a significant impact on the transient stability of the VSG and the SG \cite{liu2017pv, ying2017inertia}. In VSG-SG hybrid systems, it is discovered that decreasing the virtual inertia of the VSG can improve the transient stability \cite{zhao2025transient}. Then, a variable-inertia control method is designed to enhance the stability. However, the influence of the inertia on the transient synchronization stability is still unclear and needs further exploration.

To fill this gap, this paper investigates the inertia matching mechanisms in transient synchronization stability of hybrid systems with VSGs and SGs co-existing. The reduced relative swing equation model is built for stability analysis. Based on the model, stability mechanisms are studied from both static and dynamic perspectives. A stability index is proposed for stability comparison and mechanism elucidation. Accordingly, two inertia matching principles are discovered. Finally, a novel control strategy based on inertia matching and virtual impedance adjustment is proposed for synchronization stability enhancement and current limiting. The main contributions of this paper are summarized as follows.

1) A reduced-order relative swing equation model is derived for synchronization stability analysis in hybrid systems with VSGs and SGs.

2) Both static and dynamic synchronization stability are studied to derive a stability level index for mechanism unveiling.

3) Two inertia matching principles are elucidated. First, the inertia matching between the VSG and the SG is found to determine the transient synchronization stability. Second, the influence of the VSG share on synchronization stability is demonstrated to be significantly influenced by the matching between the VSG's inertia level and its voltage strength.

4) A strategy based on inertia matching and virtual impedance adjustment is designed to both enhance the transient synchronization stability and suppress the fault current.

The remaining sections are organized as follows: Section II developed a reduced model for a hybrid VSG-SG two-machine system. In Section III, the stability performance is analyzed, and the stability index is put forward. The two inertia matching principles are discovered. Section IV designs the control strategy. Section V conducts simulations on a two-machine system and the IEEE 39-bus system to verify the correctness of the mechanisms and effectiveness of the strategy. Section VI concludes this paper.

\section{System Modeling}

\begin{figure}[b]
\centering
\includegraphics[width=88mm]{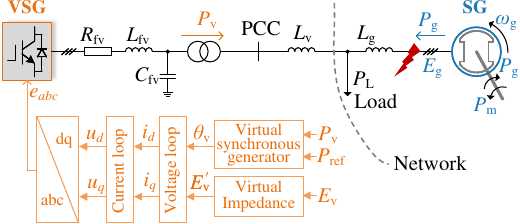}
\caption{Topology of the hybrid system under grid faults.}
\label{fig.1}
\end{figure}

\begin{figure}[b]
\centering
\includegraphics[width=88mm]{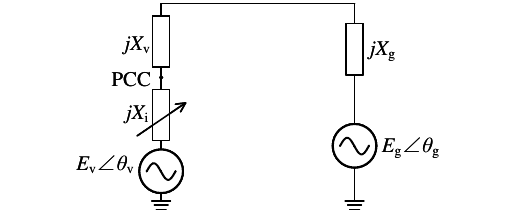}
\caption{Simplified circuit diagram of the hybrid VSG-SG system.}
\label{fig.2}
\end{figure}

To investigate the transient synchronization stability of the hybrid system with VSG and SG co-existing, a simplified two-machine hybrid system, as shown in Fig.\ref{fig.1}, is studied in this paper. It is assumed that a three-phase fault occurs in the network, leading to a network voltage dip. During the grid faults, the VSG adopts the virtual impedance control to limit the fault current. Therefore, both the VSG and the SG exhibit a voltage-source output characteristic. Then the simplified circuit diagram of the hybrid system is shown in Fig.\ref{fig.2}.

\subsection{Dynamics of VSG}
The control structure is shown in Fig.\ref{fig.1} in orange. Because the bandwidths of the voltage and current loops are much larger than the outer synchronization and virtual impedance loops, the dynamics of the voltage and current loops are ignored. Therefore, the dynamics of the VSG are modelled as follows.

\begin{subequations}\label{eq:VSG}
\begin{align}
& \frac{{{\rm{d}}{\theta _{{\rm{v}}}}}}{{{\rm{d}}t}} = {\Omega _{{\rm{ref}}}}\Delta \omega_{\rm{v}}\label{eq:VSG-a}\\
& 2{H_{{\rm{v}}}}\frac{{{\rm{d}}\Delta \omega_{\rm{v}} }}{{{\rm{d}}t}} = {P_{{\rm{vref}}}} - P_{\rm{v}} - {D_{{\rm{v}}}}\Delta \omega_{\rm{v}} \label{eq:VSG-b}
\end{align}
\end{subequations}
where ${\theta _{{\rm{v}}}}$ and $\Delta \omega_{\rm{v}} = \omega_{\rm{v}}-\omega_{0}$ are the angle and frequency deviation (in pu) of the VSG, respectively. $\omega_{\rm{v}}$ is the frequency (in pu) of the VSG and $\omega_{0} = 1.0$ pu is the nominal frequency. $\Omega_{\rm{ref}}$ is the reference electrical angular velocity (i.e., $\Omega_{\rm{ref}} = 100\pi$ rad/s). ${H_{{\rm{v}}}}$ and ${D_{{\rm{v}}}}$ are the inertia and damping of the VSG, respectively. $P_{\rm{vref}}$ and $P_{\rm{v}}$ are the reference active power and output active power of the VSG (in pu), respectively. The expression of the VSG's output active power is as follows \cite{kundur2007power}.

\begin{equation}\label{eq:Pv}
P_{\rm{v}} = \frac{E_{\rm{v}}E_{\rm{g}}\sin \left( {\theta _{{\rm{v}}}-\theta_{\rm{g}}} \right)}{X_{\rm{i}}+X_{\rm{v}}+X_{\rm{g}}}
\end{equation}
where $X_{\rm{i}}$ and $E_{\rm{v}}$ are the virtual impedance and internal voltage of the VSG, respectively. $E_{\rm{g}}$ is the SG voltage. $X_{\rm{v}}$ and $X_{\rm{g}}$ are the line impedance of the VSG and the SG, respectively. $\theta_{\rm{g}}$ is the electrical angle of the SG.

\subsection{Dynamics of SG}
The SG's dynamics under the grid fault are usually represented by the second-order swing equation. That is, 
\begin{subequations}\label{eq:SG}
\begin{align}
& \frac{{{\rm{d}}{\theta _{{\rm{g}}}}}}{{{\rm{d}}t}} = {\Omega _{{\rm{ref}}}}\Delta \omega_{\rm{g}}\label{eq:SG-a}\\
& 2{H_{{\rm{g}}}}\frac{{{\rm{d}}\Delta \omega_{\rm{g}} }}{{{\rm{d}}t}} = {P_{{\rm{m}}}} - P_{\rm{g}} - {D_{{\rm{g}}}}\Delta \omega_{\rm{g}} \label{eq:SG-b}
\end{align}
\end{subequations}
where $\Delta \omega_{\rm{g}} = \omega_{\rm{g}}-\omega_{0}$ is the frequency deviation (in pu) of the SG and $\omega_{\rm{g}}$ is the SG frequency. ${H_{{\rm{g}}}}$, ${D_{{\rm{g}}}}$, $P_{\rm{m}}$, and $P_{\rm{g}}$ are the inertia, damping, mechanical power, and electromagnetic power of the SG, respectively. The power balance principle indicates that

\begin{equation}\label{eq:Pg}
P_{\rm{g}} =  P_{\rm{L}}-P_{\rm{v}}
\end{equation}

Assuming that the load is resistive. The load consumption power is derived as follows.
\begin{equation}\label{eq:PL}
P_{\rm{L}} =  \frac{E^2_{\rm{g}}}{R_{\rm{L}}}
\end{equation}

\subsection{Synchronization Dynamics of Hybrid System}
Supposing that the ratio of the damping to the inertia is the same for the VSG and the SG. That is,
\begin{equation}\label{eq:D_H_ratio}
\frac{D_{\rm{v}}}{H_{\rm{v}}} =  \frac{D_{\rm{g}}}{H_{\rm{g}}}
\end{equation}

Subtract \eqref{eq:SG-a} from \eqref{eq:VSG-a}, and subtract \eqref{eq:SG-b} from \eqref{eq:VSG-b}. This yields a second-order \textbf{relative swing equation model} \cite{ref12}. 

\begin{subequations}\label{eq:VSG-SG}
\begin{align}
& \frac{{{\rm{d}}{\delta _{{\rm{vg}}}}}}{{{\rm{d}}t}} = {\Omega _{{\rm{ref}}}}\Delta \omega_{\rm{vg}}\label{eq:VSG-SGa}\\
& 2{H_{{\rm{vg}}}}\frac{{{\rm{d}}\Delta \omega_{\rm{vg}} }}{{{\rm{d}}t}} = {P_{{\rm{syn,ref}}}} - {P_{{\rm{syn,max}}}}\sin {\delta _{{\rm{vg}}}} - {D_{{\rm{vg}}}}\Delta \omega_{\rm{vg}} \label{eq:VSG-SGb}
\end{align}
\end{subequations}
where $\delta_{\rm{vg}} = \theta_{\rm{v}} - \theta_{\rm{g}}$ and $\Delta \omega_{\rm{vg}} = \omega_{\rm{v}} - \omega_{\rm{g}}$ are the synchronization angle and synchronization frequency between the VSG and the SG, respectively. $H_{\rm{vg}}$, $D_{\rm{vg}}$, $P_{\rm{syn,ref}}$, and $P_{\rm{syn,max}}$ are synchronization inertia, synchronization damping, reference synchronization power, and maximum output synchronization power, respectively.

\begin{subequations}\label{explanation}
\begin{align}
& {H_{{\rm{vg}}}} = \frac{{{H_{\rm{v}}}{H_{\rm{g}}}}}{{{H_{\rm{v}}} + {H_{\rm{g}}}}} \label{explanation.a}\\
& {D_{{\rm{vg}}}} = \frac{{{H_{\rm{g}}}}}{{{H_{\rm{v}}} + {H_{\rm{g}}}}}{D_{\rm{v}}} \label{explanation.b}\\
& {P_{{\rm{syn,ref}}}} = \frac{{{H_{\rm{g}}}}}{{{H_{\rm{v}}} + {H_{\rm{g}}}}}{P_{{\rm{vref}}}} - \frac{{{H_{\rm{v}}}}}{{{H_{\rm{v}}} + {H_{\rm{g}}}}} \left( {P_{{\rm{m}}}-P_{\rm{L}}} \right) \label{explanation.c} \\
& {P_{{\rm{syn,max}}}} = \frac{{{E_{\rm{g}}}{E_{\rm{v}}}}}{{{X_{\rm{i}}+X_{\rm{v}}+X_{\rm{g}}} }} \label{explanation.d}
\end{align}
\end{subequations}

The model illustrates that the synchronization dynamics between the VSG and the SG have the same form as the SG. This is because the control of the VSG essentially mimics the rotor swing dynamics of SG. The difference lies in the fact that VSG's parameters are controllable variables, whereas SG's parameters are fixed, which leads to flexibility and complexity. For the purposes of simplification, $P_{\rm{net}}$ is donated as the net input power of the SG, which is the residual power that the SG can supply after subtracting load consumption. $X_{\rm{sum}}$ represents the total impedance between the VSG's internal voltage and the SG's voltage.

\begin{subequations}\label{donation}
\begin{align}
& {P_{{\rm{net}}}} = {P_{{\rm{m}}}-P_{\rm{L}}}  \label{donation.a} \\
& X_{\rm{sum}} = {{{X_{\rm{i}}+X_{\rm{v}}+X_{\rm{g}}} }} \label{donation.b}
\end{align}
\end{subequations}

\section{Transient Synchronization Stability Mechanism Analysis}
The transient synchronization stability is analyzed from two aspects: the existence of the stable equilibrium point (SEP) (static stability) and the dynamic performance.
\subsection{Static Stability Analysis}
According to the model \eqref{eq:VSG-SG}, the condition for the existence of an SEP of transient synchronization stability is derived as,

\begin{equation}
\label{SEP_condition}
\left| {{P_{{\rm{syn,ref}}}}} \right| < {P_{{\rm{syn,max}}}}
\end{equation}

This indicates that whether the output synchronization power could cover the reference synchronization power determines the existence of an SEP.

\subsection{Dynamic Stability Analysis}

\begin{figure}[b]
\centering
\includegraphics[width=88mm]{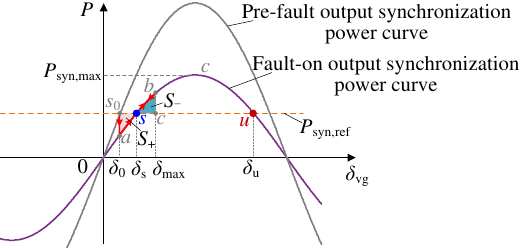}
\caption{Modified equal area criterion.}
\label{fig.4}
\end{figure}

The dynamic performance is explored using the modified equal area criterion (MEAC), as shown in Fig.\ref{fig.4}. The acceleration and deceleration areas are enclosed by the reference synchronization power curve ($P_{\rm{syn,ref}}$) and the output synchronization power curve ($P_{\rm{syn,max}}\sin\delta_{\rm{vg}}$).

\subsubsection{Dynamic Process Analysis} The dynamic process can be divided into four stages:

\textit{Stage I}: $s_0 \rightarrow a$. Before the grid fault, the pre-fault output synchronization power curve is shown as the gray solid line. The system keeps stable and the synchronization angle stays at the SEP $s_0$ of $\delta_{\rm{vg}}=\delta_0$ before the fault.

\textit{Stage II}: $a \rightarrow s$. After the fault occurs, the SG's voltage gets smaller ($E_{\rm{g}}\downarrow$) and the VSG's virtual impedance is activated ($X_{\rm{i}}>0$), leading to a reduced output power based on \eqref{explanation.c}. The fault-on output synchronization power curve is plotted with the purple solid line. Therefore, the initial point moves from point $s_0$ to point $a$. Since the reference synchronization power is larger than the output synchronization power, the synchronization frequency and the synchronization angle begin to increase according to \eqref{eq:VSG-SG}. The synchronization angle goes along with the red solid line from point $a$ to point $s$, as the red arrow shown. When the synchronization angle reaches $\delta_{\rm{s}}$ at point $s$, the reference synchronization power is equal to the output synchronization power, and the synchronization frequency reaches the maximum value. In this stage, the synchronization frequency accelerates. The increased synchronization frequency value is determined by the acceleration area closed by $s_0-a-s-s_0$, which is donated as $S_{\rm{+}}$ and painted with red in Fig.\ref{fig.4}.

\begin{equation}
\label{acceleration_area}
{S_ + }\mid_{\delta _0}^{\delta _{\text{s}}} = H_{\rm{vg}} \Delta \omega^2_{\rm{vg,s}} = \int_{{\delta _0}}^{{\delta _{\rm{s}}}} {\left( {{P_{{\rm{syn,ref}}}} - {P_{{\rm{syn,max}}}}\sin  {\delta _{{\rm{vg}}}}} \right)d{\delta _{{\rm{vg}}}}} 
\end{equation}

\textit{Stage III}: $s \rightarrow b$. After that, the synchronization angle continues to increase because the synchronization frequency is larger than zero. While the synchronization frequency, however, begins to decrease. By the time the synchronization frequency dropped to zero, the synchronization angle ceased to increase, reaching its maximum value of $\delta_{\rm{max}}$ at point $b$. In this stage, the synchronization frequency decreases. The change value can also be calculated by the deceleration area enclosed by $s-b-c$, which is donated as $S_{\rm{-}}$ and painted with blue in Fig.\ref{fig.4}.

\begin{equation}
\label{deceleration_area}
\begin{aligned} 
{S_-}\mid_{\delta _{\text{s}}}^{\delta _{\text{max}}} &= {H_{\rm{vg}}}\left( \Delta \omega_{\rm{vg,s}}^2 - \Delta \omega_{\rm{vg,b}}^2 \right)\\
      &= \int_{\delta_s}^{\delta_{\rm{max}}} \left( P_{\rm{syn,max}}\sin \delta_{\rm{vg}} - P_{\rm{syn,ref}} \right)d\delta_{\rm{vg}}
\end{aligned}
\end{equation}

\textit{Stage IV}: $b \rightarrow s$ or $b \rightarrow u \rightarrow$. Subsequent dynamics can be categorized into stable and unstable patterns.

(1) Stable pattern: $\delta_{\rm{max}}<\delta_{\rm{u}}$. If the synchronization frequency decreases to zero before the synchronization angle reaches to the UPE ($\delta_{\rm{max}}<\delta_{\rm{u}}$), the maximum angle point $b$ is located before the UEP $u$. At point $b$, the reference synchronization power is smaller than the output synchronization power. The synchronization frequency will then decrease starting from zero, thereby reducing the synchronization angle. Then the synchronization angle returns to the SEP during the fault-on period. Based on the analysis, the synchronization frequency increment from point $a$ to point $s$ must be smaller than the synchronization frequency decrement from point $s$ to UEP $u$ in this pattern. That is,

\begin{equation}
\label{Stable_condition}
S_+\mid_{\delta _0}^{\delta _{\text{s}}}<S_-\mid_{\delta _0}^{\delta _{\text{u}}} 
\end{equation}

(2) Unstable pattern: $\delta_{\rm{max}}>\delta_{\rm{u}}$. However, if the synchronization angle increases to the UEP $u \left(\delta_{\rm{u}}\right)$ before the synchronization frequency decrease to zero, both the synchronization frequency and synchronization angle will continue to increase after passing through UEP, leading to LOS. In this pattern, the acceleration area is larger than the deceleration area.

\begin{equation}
\label{Unstable_condition}
S_+\mid_{\delta _0}^{\delta _{\text{s}}}>S_-\mid_{\delta _0}^{\delta _{\text{u}}} 
\end{equation}

\subsubsection{Key Factors Analysis} 
\begin{figure}[h]
\centering
\includegraphics[width=88mm]{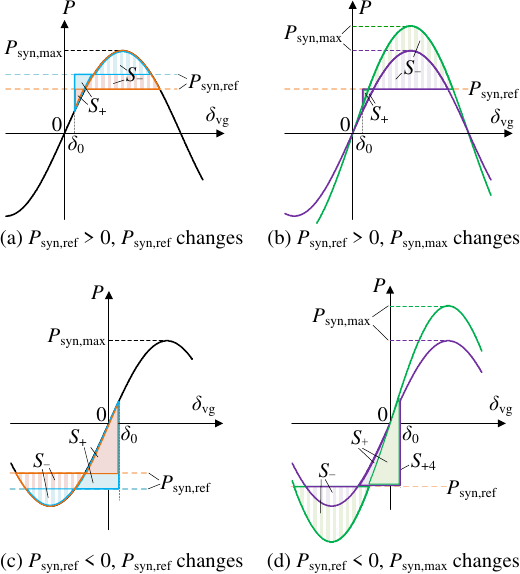}
\caption{Synchronization power influence analysis.}
\label{fig.5}
\end{figure}

Based on the analysis above, it is found that the relationship between the reference synchronization power and the maximum output synchronization power determines the acceleration/deceleration area and the synchronization stability. The influence of the synchronization powers on the dynamic stability is shown in Fig.\ref{fig.5}.

(1) When $P_{\rm{syn,ref}}>0$, the influences of the reference synchronization power and output synchronization power on the acceleration/deceleration area are shown in Fig.\ref{fig.5}(a)-(b). It shows that when the synchronization reference power gets larger ($P_{\rm{syn,ref}} \uparrow$), the acceleration area becomes larger ($S_+\mid_{\delta _0}^{\delta _{\text{s}}} \uparrow$), and the deceleration area gets smaller ($S_-\mid_{\delta _0}^{\delta _{\text{u}}}  \downarrow$). Based on \eqref{Stable_condition} and \eqref{Unstable_condition}, an unstable pattern is more likely to happen, and the system's synchronization stability deteriorates. How the output synchronization power ($P_{\rm{syn,max}}$) affects the dynamic performance is shown in Fig.\ref{fig.5}(b). As $P_{\rm{syn,max}}$ gets larger, $S_+\mid_{\delta _0}^{\delta _{\text{s}}}$ decreases and $S_-\mid_{\delta _0}^{\delta _{\text{u}}}$ increases, leading to improved synchronization stability.

(2) When $P_{\rm{syn,ref}}<0$, the influences of the reference synchronization power and output synchronization power on the acceleration/deceleration area are shown in Fig.\ref{fig.5}(c)-(d). It shows that when the reference synchronization power increases ($P_{\rm{syn,ref}} \uparrow$) or the output synchronization power gets larger ($P_{\rm{syn,max}} \uparrow$), $S_+\mid_{\delta _0}^{\delta _{\text{s}}} $ becomes smaller and $S_-\mid_{\delta _0}^{\delta _{\text{u}}} $ gets larger. The stable pattern of \eqref{Stable_condition} is more likely to occur. Therefore, synchronization stability is improved.

In all, smaller $|{{P_{{\rm{syn,ref}}}}}|$ and larger ${P_{{\rm{syn,max}}}}$ lead to better dynamics performance, which is beneficial to the transient synchronization stability.

\subsection{Proposed Stability Level Index}
Based on the static stability analysis, the magnitude relationship between the output synchronization power and the reference synchronization power determines whether an equilibrium point exists. Besides, larger output synchronization power and smaller reference synchronization power improve the dynamic stability. Therefore, the stability level index is proposed as follows.

\begin{empheq}[box=\fbox]{equation}
\label{margin_index}
\lambda  = 1 - \frac{{\left| {{P_{{\rm{syn,ref}}}}} \right|}}{{{P_{{\rm{syn,max}}}}}}
\end{empheq}

(1) If {$\lambda<0$}, the system will lose synchronization for no SEP.

(2) If {$\lambda =0$}, there is an unstable equilibrium point. Therefore, LOS will also occur.

(3) If {$0<\lambda<1$}, there is definitely an SEP existing based on \eqref{SEP_condition} and \eqref{margin_index}. When $\lambda$ gets larger, the dynamic performance is improved, and the synchronization stability will be enhanced.

(4) If $\lambda=1$, the synchronization stability reaches its best in this case. The system will definitely converge to an SEP under the effect of damping.

As is known, SCR is widely used to represent the strength of the grid to which the VSG is connected. SCR is defined as the ratio between the short-circuit power of the grid and the rated power of the converter. It is usually calculated by the inverse of the line impedance between the converter and the grid as follows \cite{he2021pll}.

\begin{equation}
\label{SCR}
\gamma_{\rm{SCR}} =  \frac{1}{|Z|} = \frac{1}{X_{\rm{g}}+X_{\rm{v}}}
\end{equation}

According to \eqref{explanation.c} and \eqref{margin_index}, the relationship between the SCR and the proposed index is derived.

\begin{equation}
\label{SCR_and_index}
\lambda  = 1 - \left( \frac{1}{\gamma_{\rm{SCR}}} + X_{\rm{i}} \right )\frac{{\left| {{P_{{\rm{syn,ref}}}}} \right|}}{{{E_{{\rm{v}}}}}{{E_{{\rm{g}}}}}}
\end{equation}

It can be seen that as the grid strength between the VSG and the SG increasing, the maximum output capability of the VSG is promoted, leading to improved transient stability. The conclusion is consistent with the previous work \cite{li2024transient}. Therefore, SCR and the proposed index can both reflect the transient stability level. However, apart from the grid strength, the fault depth and the parameters of the VSG itself would also influence the stability. As a result, the proposed index enables a more comprehensive assessment of transient stability and the influencing factors than SCR.

It is important to note that the proposed index is intended to serve as an indicator for the comparison of stability levels and the underlying mechanism, rather than as a tool for evaluating stability.

\subsection{Inertia Matching Mechanism Between VSG and SG}
\subsubsection{Existence of SEP}
Based on \eqref{eq:VSG-SG}-\eqref{donation}, the condition for the existence of an SEP ($\lambda>0$) is derived as,
\begin{equation}
\label{SEP_condition2}
\left| {\frac{{{P_{{\rm{vref}}}}}}{{{H_{\rm{v}}}}} - \frac{{{P_{{\rm{net}}}}}}{{{H_{\rm{g}}}}}} \right| < \frac{{{E_{\rm{g}}}{E_{\rm{v}}}\left( {{H_{\rm{v}}} + {H_{\rm{g}}}} \right)}}{{{H_{\rm{v}}}{H_{\rm{g}}} {{X_{\rm{sum}}} }}}
\end{equation}

(1) When ${H_{\rm{v}}}/{H_{\rm{g}}}>{P_{\rm{vref}}}/{P_{\rm{net}}}$, \eqref{SEP_condition2} is transformed as,
\begin{equation}
\label{SEP_solution3-1}
\left( {{P_{{\rm{net}}}} - \frac{{{E_{\rm{g}}}{E_{\rm{v}}}}}{{{X_{{\rm{sum}}}}}}} \right)\frac{{{H_{\rm{v}}}}}{{{H_{\rm{g}}}}} < \frac{{{E_{\rm{g}}}{E_{\rm{v}}}}}{{{X_{{\rm{sum}}}}}} + {P_{{\rm{vref}}}}
\end{equation}

a) If $E_{\rm{g}}>{{{P_{\rm{net}}}{X_{\rm{sum}}}}}/{{{E_{{\rm{V}}}}}}$, \eqref{SEP_solution3-1} is always satisfied. This means that the SEP exists when the inertia of the VSG and the voltage of the SG is significant.

b) If $E_{\rm{g}}<{{{P_{\rm{net}}}{X_{\rm{sum}}}}}/{{{E_{{\rm{V}}}}}}$, \eqref{SEP_solution3-1} is calculated as ${H_{\rm{v}}}/{H_{\rm{g}}}<\left( {E_{\rm{g}}E_{\rm{v}}+P_{\rm{vref}}X_{\rm{sum}}} \right) / \left( {P_{\rm{net}}X_{\rm{sum}}-E_{\rm{g}}E_{\rm{v}}} \right)$. This indicates that when the fault is severe, the inertia of the VSG can't be too large. Otherwise, LOS will occur for no SEP.

(2) When ${H_{\rm{v}}}/{H_{\rm{g}}}<{P_{\rm{vref}}}/{P_{\rm{net}}}$, \eqref{SEP_condition2} is derived as,
\begin{equation}
\label{SEP_solution3-1}
\left( {{P_{{\rm{net}}}} - \frac{{{E_{\rm{g}}}{E_{\rm{v}}}}}{{{X_{{\rm{sum}}}}}}} \right)\frac{{{H_{\rm{v}}}}}{{{H_{\rm{g}}}}} > \frac{{{E_{\rm{g}}}{E_{\rm{v}}}}}{{{X_{{\rm{sum}}}}}} + {P_{{\rm{vref}}}}
\end{equation}

a) If $E_{\rm{g}}>{{{P_{\rm{net}}}{X_{\rm{sum}}}}}/{{{E_{{\rm{V}}}}}}$, \eqref{SEP_solution3-1} is never true. This means that small VSG's inertia and large SG's voltage lead to no SEP.

b) If $E_{\rm{g}}<{{{P_{\rm{net}}}{X_{\rm{sum}}}}}/{{{E_{{\rm{V}}}}}}$, \eqref{SEP_solution3-1} is calculated as ${H_{\rm{v}}}/{H_{\rm{g}}}> \left( {E_{\rm{g}}E_{\rm{v}}+P_{\rm{vref}}X_{\rm{sum}}} \right) /\left( {P_{\rm{net}}X_{\rm{sum}}-E_{\rm{g}}E_{\rm{v}}} \right)$. This indicates that when the fault is severe, the inertia of the VSG can't be too small.

\begin{figure}[h]
\centering
\includegraphics[width=88mm]{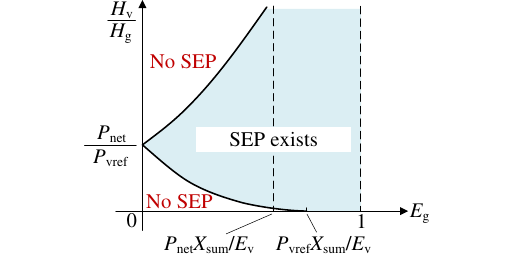}
\caption{Range of the inertia matching constant to guarantee the SEP.}
\label{fig.3}
\end{figure}

In summary, the solution for \eqref{SEP_condition} is then calculated based on different voltage dip levels, as shown in Fig.\ref{fig.3}. \textbf{When there is a severe fault leading to a low voltage, either too large or too small inertia matching between the VSG and the SG would lead to LOS for no SEP}.

\subsubsection{Stability Level Comparison}
When there exists an SEP, the stability level index is rewritten based on \eqref{explanation}.
\begin{equation}
\label{margin_index2}
\lambda  = 1 - \frac{{\left| {\frac{{{H_{\rm{g}}}}}{{{H_{\rm{v}}} + {H_{\rm{g}}}}}{P_{{\rm{vref}}}} - \frac{{{H_{\rm{v}}}}}{{{H_{\rm{v}}} + {H_{\rm{g}}}}}{P_{{\rm{net}}}}} \right|}}{{\frac{{{E_{\rm{g}}}{E_{\rm{v}}}}}{{{X_{\rm{sum}}}}}}}
\end{equation}

The partial derivative of $\lambda$ with respect to the inertia matching constant between the VSG and the SG (${H_{\rm{v}}}/H{\rm{g}}$) is then calculated.

(1) When $H_{\rm{v}}/H_{\rm{g}}<P_{\rm{ref}}/P_{\rm{net}}$, the partial derivative is derived as,
\begin{equation}
\label{derivative1}
\frac{{\partial \lambda }}{{\partial \left( {H_{\rm{v}}}/H_{\rm{g}}\right)}} =  \frac{{{H^2_{\rm{g}}}\left( {{P_{{\rm{vref}}}} + {P_{{\rm{net}}}}} \right)}}{{\frac{{{E_{\rm{g}}}{E_{\rm{v}}}}}{{{X_{\rm{sum}}}}}{{\left( {{H_{\rm{v}}} + {H_{\rm{g}}}} \right)}^2}}}\
\end{equation}

\begin{figure}[t]
\centering
\includegraphics[width=88mm]{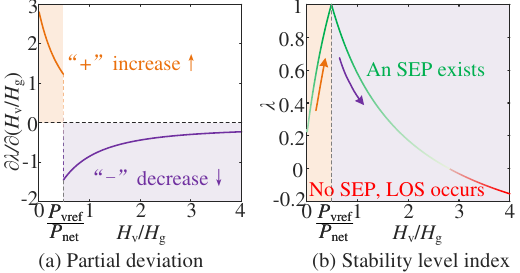}
\caption{Influence of the inertia matching on the synchronization stability.}
\label{fig.6}
\end{figure}

It can be seen that ${{\partial \lambda }}/{{\partial \left( {H_{\rm{v}}}/H_{\rm{g}}\right)}}>0$. This indicates that the synchronization stability level is improved when the inertia matching constant increases in this range. This trend is shown with yellow areas in Fig.\ref{fig.6}(a)-(b).

(2) When $H_{\rm{v}}/H_{\rm{g}}>P_{\rm{ref}}/P_{\rm{net}}$, the partial derivative is derived as,
\begin{equation}
\label{derivative2}
\frac{{\partial \lambda }}{{\partial \left( {H_{\rm{v}}}/H_{\rm{g}}\right)}} =  -\frac{{{H^2_{\rm{g}}}\left( {{P_{{\rm{vref}}}} + {P_{{\rm{net}}}}} \right)}}{{\frac{{{E_{\rm{g}}}{E_{\rm{v}}}}}{{{X_{\rm{sum}}}}}{{\left( {{H_{\rm{v}}} + {H_{\rm{g}}}} \right)}^2}}}\
\end{equation}

Obviously that ${{\partial \lambda }}/{{\partial \left( {H_{\rm{v}}}/H_{\rm{g}}\right)}}<0$. This shows that the synchronization stability deteriorates when the inertia matching constant increases when $H_{\rm{v}}/H_{\rm{g}}>P_{\rm{ref}}/P_{\rm{net}}$. This is shown in the purple areas in Fig.\ref{fig.6}.

A conclusion can be drawn that when the inertia of the VSG is matched with that of the SG, the system achieves optimal stability. The physical interpretation of this phenomenon is as follows. The inertia constant reflects the inertia-response power to the frequency disturbance of the VSG/SG itself. $P_{\rm{vref}}$ and $P_{\rm{net}}$ can be regarded as the portions of generation power that the VSG and the SG are required to undertake during a disturbance, respectively. \textbf{For two units whose inertia constants are proportional to their assigned generation power, the corresponding inertial response power is also proportional to the power they undertake. Under power disturbances and inertial response, their frequency dynamics exhibit coherent characteristics. Therefore, their frequencies remain nearly identical, and the angle difference between them remain unchanged. The system achieves optimal transient synchronization stability performance.}

The partial derivative and the index with different inertia matching constants are drawn in Fig.\ref{fig.6}. The results are fully consistent with the preceding analysis.

\subsection{Inertia-Voltage Strength Matching Mechanism of VSG}
To simplify the analysis, it is assumed that the VSG and the SG operate at the same power factor of $\cos\phi$ under pre-fault steady-state conditions. That is,
\begin{equation}
\label{ratio_VSG_SG}
\frac{P_{\text{vref}}}{S_{\text{v}}}=\frac{P_{\text{m}}}{S_{\text{g}}}=\cos \phi 
\end{equation}

In practice, to maintain consistent voltage drops along the lines, the line impedance of a generator is typically set to be inversely proportional to its rated capacity. This also applies to the configuration of the virtual impedance. And the inertia constant of the VSG/SG is proportional to the rated capacity of the VSG/SG itself. That is,
\begin{subequations}
\label{Xv_VSG}
\begin{align}
    X_{\mathrm{v}} S_{\mathrm{v}} &= a X_{\mathrm{g}} S_{\mathrm{g}} \label{Xv} \\
    X_{\mathrm{i}} S_{\mathrm{v}} &= b X_{\mathrm{g}} S_{\mathrm{g}} \label{Xi} \\
    \frac{H_{\mathrm{v}}}{S_{\mathrm{v}}} &= c \frac{H_{\mathrm{g}}}{S_{\mathrm{g}}} \label{Hv}
\end{align}
\end{subequations}
where $a$ is the ratio of line voltage drop per unit length between VSG and SG. $b$ is the ratio of line voltage drop per unit length between the VSG's virtual impedance and the SG's line. $c$ is the inertia level ratio between the VSG and the SG. If $a=1$, it means that the voltage drop per unit length along the VSG's and the SG's lines are the same. Then, $1/(a+b)$ can be regarded as the voltage strength of the VSG. A larger $1/(a+b)$ indicates a smaller equivalent output impedance of the VSG and a smaller voltage drop between the internal voltage of the VSG and the PCC, implying a stronger interconnection strength between the VSG and the grid.

Denote $\eta$ as the capability ratio between the VSG and the SG. It can reflect the VSG's penetration.
\begin{equation}
\label{eta_defination}
\eta = \frac{S_{\rm{v}}}{S_{\rm{g}}}
\end{equation}

Accordingly, the partial derivative of $\lambda$ with respect to the capacity ratio between VSG and SG is derived.
\begin{equation}
\label{partial_derivative_ratio}
\frac{\partial \lambda}{\partial \eta}=-\frac{X_{\text{g}}\left| \left( 1-c \right) \cos \phi S_{\text{g}}+cP_{\text{L}} \right|}{E_{\text{g}}E_{\text{v}}\left( c\eta +1 \right) ^2}\left( 1-c\left( a+b \right) \right) 
\end{equation}

And the conclusion is concluded that,
\begin{equation}
\label{deta-sign}
\frac{\partial \lambda}{\partial \eta}\left\{ \begin{array}{c}
	>0,\text{if}\;{c}>{{1}/{a+b}}\\
	<0,\text{if}\;{c}<{{1}/{a+b}}\\
\end{array} \right.
\end{equation}

\begin{figure}[t]
\centering
\includegraphics[width=88mm]{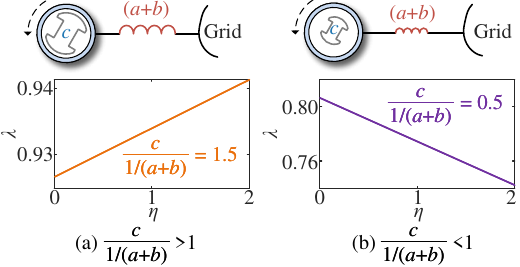}
\caption{Influence of the matching between inertia and strength of the VSG on the stability.}
\label{fig.6-1}
\end{figure}

This shows that \textbf{the influence of the VSG's penetration on the synchronization stability is greatly influenced by the relationship between its inertia level and voltage strength}: in circumstances where the inertial level of the VSG is comparatively elevated in relation to its voltage strength, the deployment of more VSGs is advantageous in terms of enhancing stability, as shown in Fig.\ref{fig.6-1}(a). Cutting out VSGs will worsen system stability. While in situations where the inertial level of the VSG is comparatively weak in relation to its voltage strength, cutting some VSGs off has a beneficial effect on the synchronization stability, as depicted in Fig.\ref{fig.6-1}(b). In power systems with high penetration of inverter-based VSGs, the switching on and off of inverters causes random variations in the VSG's proportion \cite{li2022hierarchical}. Therefore, the coordinated matching between the VSG's inertia level and its voltage strength, i.e., $c = 1/(a+b)$ is critical for ensuring robustness and stability.

\subsection{Phase Portrait Based Stability Region}
The inverse time integral-based method  \cite{chiang2002stability}
is used to draw the stability region on the phase portrait plane. When the initial point lies within the stability region, the system will be stable. Otherwise, if the initial point is outside the region, LOS will occur. Hence, the larger the stability region, the better the synchronization stability performance.


\begin{figure}[b]
\centering
\includegraphics[width=88mm]{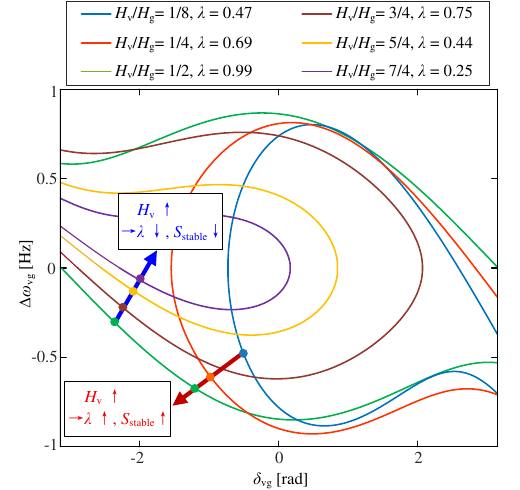}
\caption{The stability regions under different inertia matching constants.}
\label{fig.8}
\end{figure}

The stability regions with different inertia matching instant are drawn in Fig.\ref{fig.8}. When the inertia matching increases from 1/8 to 1/2, the stability region enlarges, as shown with the solid red arrow. And the proposed stability level index also increases. While when the inertia matching increases from 1/2 to 7/4, the stability region shrinks and the stability level index decreases. Two main conclusion are given from Fig.\ref{fig.8}.

(1) Greater inertia of the VSG does not necessarily lead to better synchronization stability. With the increase of the inertia matching constant, the stability region first expands and then contracts.

(2) When the inertia matching constant between the VSG and the SG is equal to the power ratio, i.e., $H_{\rm{v}}/H{\rm{g}} = P_{\rm{vref}}/P_{\rm{net}}$ = 0.4878 in Fig.\ref{fig.6}, the synchronization stability performance is optimal.

(3) The proposed index can accurately reflect the stability performance.

\begin{figure}[h]
\centering
\includegraphics[width=88mm]{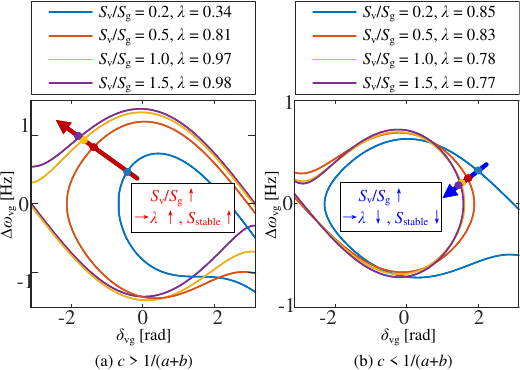}
\caption{The stability regions under different penetration of VSG when (a) $c>1/(a+)$ and (b) $c<1/(a+b)$.}
\label{fig.8-1}
\end{figure}

Moreover, the stability regions with different penetration of VSG are plotted in Fig.\ref{fig.8-1}. In Fig.\ref{fig.8-1}(a), the inertia level of the VSG is large relative to its voltage strength ($c>1/(a+b)$). The figure shows that the stability region is expanded with the increasing penetration of the VSG. The derived stability level index also increases, which is consistent with the trend. While when the inertia level of the VSG is small relative than its voltage strength ($c<1/(a+b)$), the stability region reduces as the penetration of the VSG increases, so as the stability level index. This proves the effectiveness of the proposed index and the influence of the VSG's inertia-voltage strength matching on the transient synchronization stability.

\section{Stability Control Strategy}
\subsection{Inertia Matching Control}
Based on the analysis in Section III, the inertia matching constant between the VSG and the SG determines both the static and dynamic stability of the hybrid system. The inertia of the SG is determined by the mass of its rotor and remains constant. While the virtual inertia of the VSG is governed by its control strategy and can be flexibly adjusted. As a result, the inertia matching control, which adjusts the inertia parameter of the VSG, is designed as follows.
\begin{empheq}[box=\fbox]{equation}
\label{control}
{H_{\rm{v}}}=\frac{P_{\rm{vref}}}{P_{\rm{net}}}{H_{\rm{g}}}
\end{empheq}

\subsection{Virtual Impedance Setting of VSG}
\subsubsection{Current Limitation Range}First, to restrict the fault current to protect the VSG from damage, the virtual impedance can't be too small. The current amplitude of the VSG is calculated as,
\begin{equation}
\label{current_expression}
\left| {{I_{\rm{v}}}} \right| = \left| {\frac{{{E_{\rm{v}}}{e^{j{\theta _{\rm{v}}}}} - {E_{\rm{g}}}{e^{j{\theta _{\rm{g}}}}}}}{{j {{X_{\rm{sum}}}} }}} \right| = \frac{{\left| {{E_{\rm{v}}}{e^{j{\delta _{{\rm{vg}}}}}} - {E_{\rm{g}}}} \right|}}{X_{\rm{i}} + X_{\rm{v}} + X_{\rm{g}}}
\end{equation}

Based on the analysis in Section III-B, the maximum synchronization angle is the angle at UEP to guarantee the stability. Then the maximum current amplitude of the VSG under the constraint is,
\begin{equation}
\label{current_max}
I_{\rm{vmax}} = \frac{|E_{\rm{v}}e^{j\delta_{\rm{u}}} - E_{\rm{g}}|}{X_{\rm{i}} + X_{\rm{v}} + X_{\rm{g}}} 
\end{equation}

It should meet the requirements as follows,
\begin{equation}
\label{current_limitation}
I_{\rm{vmax}} \leq   I_{\rm{lim}}
\end{equation}

If the solution of \eqref{current_limitation} is larger than zero, the range of the virtual impedance is governed by \eqref{current_limitation}. Otherwise, if the solution is smaller than zero, no virtual impedance is needed. Therefore, the virtual impedance should satisfies,
\begin{equation}
\label{virtual_impedance}
X_{\rm{i}} \geq \max\{{\frac{|E_{\rm{v}}e^{j\delta_{\rm{u}}} - E_{\rm{g}}|}{I_{\rm{lim}}} - X_{\rm{v}} - X_{\rm{g}}},0\}
\end{equation}

\subsubsection{Inertia-Strength Matching Requirement}
Based on Section III-D, the inertia-strength matching of the VSG also plays an important role in the transient stability. When the inertia level of the VSG is relatively larger than its voltage strength, the decreasing share of the VSG (e.g., VSGs' disconnection) will deteriorate the stability. While in systems with the VSG's inertia level relatively smaller than its voltage strength, increased VSG capacity (e.g., VSGs cut in) must lead to reduced system stability. As a result, the virtual impedance should be set to match the VSG's inertia to ensure reliable and robust synchronization stability.

\begin{equation}
\label{inerita_strength_matching}
    c=\frac{1}{a+b}\Rightarrow X_{\text{i}}=\frac{H_{\text{g}}}{H_{\text{v}}}X_{\text{g}}-X_{\text{v}}
\end{equation}

\begin{figure}[b]
\centering
\includegraphics[width=88mm]{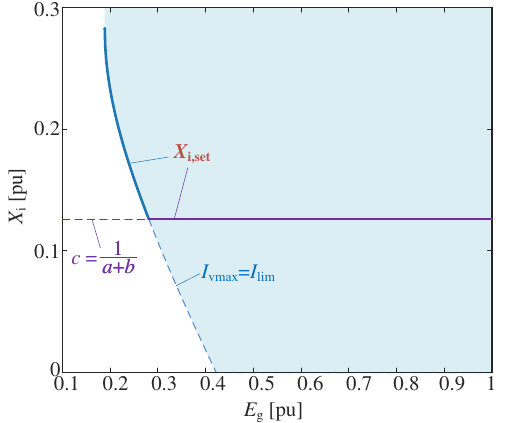}
\caption{Range of the virtual impedance to limit the fault current.}
\label{fig.9}
\end{figure}

The allowable virtual impedance of the VSG to limit its current under different grid faults is painted with blue in Fig.\ref{fig.9}. It can be seen that when the fault is shallow ($E_{\rm{g}}>0.423$ pu), there is no need to employ virtual impedance. If the fault is severe and the voltage drops below 0.423 pu, a larger virtual impedance of the VSG should be set to limit the current as the fault gets severe. Besides, the inertia-strength matching condition is drawn with a purple line in Fig.\ref{fig.9}. For a shallow grid voltage sag, the inertia–strength matching condition is able to ensure the current remains within the allowable limit. However, for deep voltage sags, the matched value of the virtual impedance cannot adequately suppress the output current. Therefore, the virtual impedance must be set to the minimum value that satisfies the current-limiting requirement.

Based on \eqref{virtual_impedance} and \eqref{inerita_strength_matching}, the virtual impedance of the VSG is set as,
\begin{empheq}[box=\fbox]{equation}
\label{virtual_impedance_set}
X_{\text{i,set}}=\max \left\{ \frac{H_{\text{g}}}{H_{\text{v}}}X_{\text{g}}-X_{\text{v}},\frac{|E_{\text{v}}e^{j\delta _{\text{u}}}-E_{\text{g}}|}{I_{\lim}}-X_{\text{v}}-X_{\text{g}} ,0\right\}
\end{empheq}

\subsection{Static Stability Analysis Under the Control}
In this case, the SEP existence condition \eqref{SEP_condition2} is always satisfied.
\begin{equation}
\label{control_SEP}
\left| {\frac{{{P_{{\rm{vref}}}}}}{{{H_{\rm{v}}}}} - \frac{{{P_{{\rm{net}}}}}}{{{H_{\rm{g}}}}}} \right| =0< \frac{{{E_{\rm{g}}}{E_{\rm{v}}}\left( {{H_{\rm{v}}} + {H_{\rm{g}}}} \right)}}{{{H_{\rm{v}}}{H_{\rm{g}}} {{X_{\rm{sum}}} }}}
\end{equation}

From Fig.\ref{fig.3}, it can be seen that only when the inertia matching satisfies \eqref{control}, the SEP exists in the whole voltage dip range. Therefore, the static stability is guaranteed even under severe grid faults.

\subsection{Dynamic Stability Analysis Under the Control}
The stability region reaches the maximum value according to the dynamic performance analysis.

\section{Simulation Verifications}
\subsection{Two-Machines Hybrid System}
Simulations of the hybrid system in Fig.\ref{fig.1} are conducted on MATLAB/SIMULINK platform. The parameters of the system are listed in the TABLE.\ref{tab:two_machine_system_params}.
\begin{table}[htbp]
  \centering 
  \caption{System Parameters} 
  \label{tab:two_machine_system_params} 
  \begin{tabular}{l c r}
    \toprule 
    \textbf{Name} & \textbf{Symbol} & \textbf{Value} \\
    \midrule 
    Rated Voltage & $U_{\rm{n}}$ & 95.22 V \\
    Rated Power & $S_{\rm{n}}$ & 1 kW \\
    Rated Frequency & $f_{\rm{n}}$ & 50 Hz \\
    Inertia of SG & $H_{\rm{g}}$ & 40 s \\
    Damping of SG & $D_{\rm{g}}$ & 20 pu \\
    Mechanical Power of SG & $P_{\rm{m}}$ & 1 kW \\
    Line Inductance of SG & $L_{\rm{g}}$ & 2.9 mH \\
    Rated Power of VSG & $S_{\rm{vsg}}$ & 1 kW \\
    Reference Power of VSG Under Fault & $P_{\rm{ref}}$ & 0.3 pu \\
    Virtual Internal Voltage of VSG & $E_{\rm{v}}$ & 1.0 pu \\
    Line Inductance of VSG & $L_{\rm{v}}$ & 9.2 mH \\
    Virtual Impedance of VSG & $L_{\rm{i}}$ & 1.45 mH \\
    Load Power & $P_{\rm{L}}$ & 1 kW \\
    Fault Voltage & $E_{\rm{gf}}$ & 0.2 pu \\
    \bottomrule 
  \end{tabular}
\end{table}

\subsubsection{Inertia Matching Mechanism}
\begin{figure}[b]
\centering
\includegraphics[width=88mm]{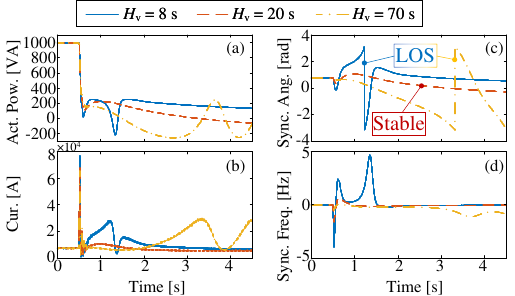}
\caption{Simulation results of the two-machines hybrid system.}
\label{fig.10}
\end{figure}
In Fig.\ref{fig.10}, the simulation results with different inertia of the VSG ($H_{\rm{v}}$) are plotted. The fault occurs at t = 0.5 s, resulting in a fault voltage of 0.2 pu. When the inertia of the VSG is small of 8 s, the synchronization frequency is always larger than zero and the synchronization angle son increases beyond $\pi$ after the fault occurs, as the solid blue line shown. Los occurs and the active power and current of the VSG oscillates. It should be mentioned that, though the synchronization frequency decreased to zero finally. The synchronization angle just converges to an SEP at the next cycle, which suffers from pole slip and is not allowed in practical engineering \cite{liu2025state}. When the inertia of the VSG is large of 70 s, the results are shown with yellow chain lines. The synchronization angle begins to decrease after the fault, leading to LOS and power oscillations finally. These indicate that LOS occurs when the virtual inertia of the VSG is either small or large. When the VSG's inertia is moderate at 20s, the system remains stable as the dotted yellow lines shown. These results are consistent with the previous conclusion of the inertia matching principle.

\subsubsection{Inertia-Voltage Strength Matching Mechanism of VSG}
The synchronization stability index (SSI) for simulation is commonly used to assess the synchronization stability\cite{tan2023debiased}. It is defined as follows.
\begin{equation}\label{eq:SSI}
\rm{SSI} = \frac{2\pi-\delta_{vg,max}}{2\pi+\delta_{\rm{vg,max}}}
\end{equation}
where $\delta_{\rm{vg,max}}$ is the maximum synchronization angle between the VSG and the grid.
\begin{figure}[b]
\centering
\includegraphics[width=88mm]{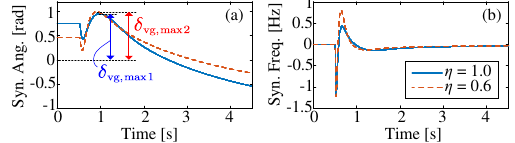}
\caption{The inertia level of the VSG is large relative to its voltage strength ($a = 2, b = 0.4$, and $c = 0.75$. i.e., $c>1/(a+b)$).}
\label{fig.11}
\end{figure}

\begin{figure}[b]
\centering
\includegraphics[width=88mm]{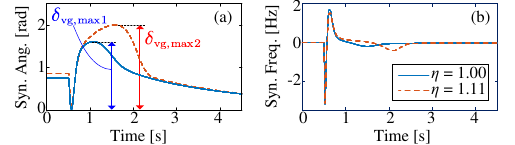}
\caption{The inertia level of the VSG is weak relative to its voltage strength ($a = 2, b = 0.4$, and $c = 0.25$. i.e., $c<1/(a+b)$).}
\label{fig.12}
\end{figure}

When the inertia level of the VSG is large relative to its voltage strength, the simulation results are shown in Fig.\ref{fig.11}. The synchronization angles with the capability ratio between the VSG and the SG of 1.0 and 0.6 are plotted in Fig.\ref{fig.11}(a) with a solid blue line and a dashed red line. It indicates that the maximum synchronization angle with $\eta = 0.6$ is larger than that with $\eta = 1.0$, i.e.,  $\delta_{\rm{vg,max2}}>\delta_{\rm{vg,max1}}$. Based on \eqref{eq:SSI}, larger $\delta_{\rm{vg,max}}$ represents a poorer stability. This shows that the increase in VSG capacity has improved the system's synchronization stability in this situation.

When the inertia level of the VSG is weak relative to its voltage strength, the simulation results are shown in Fig.\ref{fig.12}. The synchronization angles with the capability ratio between the VSG and the SG of 1.0 and 1.11 are plotted in Fig.\ref{fig.11}(a) with a solid blue line and a dashed red line. It indicates that the maximum synchronization angle with $\eta = 1.11$ is larger than that with $\eta = 1.0$. According to \eqref{eq:SSI}, synchronization stability with $\eta = 1.11$ is worsen than that with $\eta = 1.0$. This proves that the increase in VSG capacity deteriorates the synchronization stability when the inertia level of the VSG is weaker than its voltage strength.

\subsection{IEEE 39-bus System}
\begin{figure}[b]
\centering
\includegraphics[width=88mm]{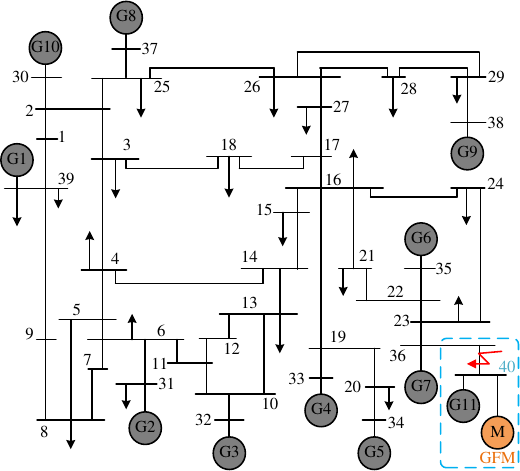}
\caption{Modified IEEE 39-bus system.}
\label{fig.13}
\end{figure}

Simulations are conducted on the IEEE 39-bus system to verify the effectiveness of the proposed strategy. The VSG and an SG (G11) are connected to the bus 36, as shown in Fig.\ref{fig.13}. The fault occurs at 0.5 s at the line between bus 36 and bus 40, where bus 40 is the point of interconnection (POI) of the GFM and the G11.

\begin{figure}[t]
\centering
\includegraphics[width=88mm]{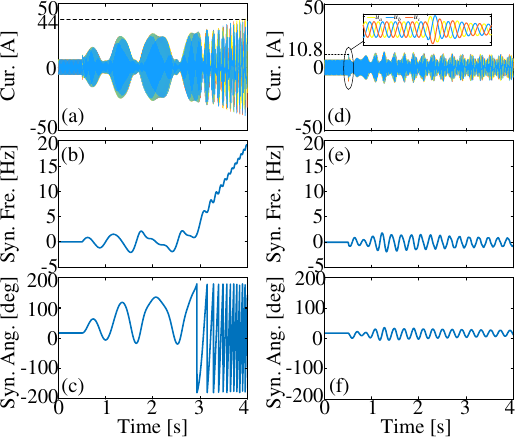}
\caption{Simulation results of the modified IEEE 39-bus system.}
\label{fig.14}
\end{figure}

The three-phase currents, synchronization frequency, and synchronization angle of the VSG with the traditional control are shown in Fig.\ref{fig.14}(a)-(c), respectively. It can be seen that he synchronization angle reaches $180^\circ$ at 2.9 s, as shown in Fig.\ref{fig.14}(c). This indicates that the GFM loses synchronization with the main grid. The maximum current reaches 44 A = 7.23$I_{\rm{n}}$, which has greatly exceeded the VSG's allowable current range of $I_{\rm{max}} = 1.8 I_{\rm{n}}$. After adopting the proposed control strategy, the three-phase currents, synchronization frequency, and synchronization angle of the VSG are shown in Fig.\ref{fig.14}(d)-(f), respectively. The inertia of the VSG is adjusted to 0.75 s in this case, and the virtual impedance of the VSG is also set to $j$0.1333 pu. The result in Fig.\ref{fig.14}(f) shows that the synchronization angle converges, indicating that the transient synchronization stability is maintained. Besides, the maximum fault current is 10.8 A, which is within the allowable maximum current of 10.9 A, as shown in Fig.\ref{fig.14}(d). This proves that the proposed strategy can both improve the transient synchronization stability and suppress the fault current.

\section{Conclusion}
The future power system is expected to feature a coexistence of inverter-based VSGs and rotor-dominated SGs. Therefore, the transient synchronization stability of the hybrid system consisting of VSGs and SGs is studied in this paper. A relative swing equation model is derived for a two-machine VSG-SG system. Synchronization stability is analyzed from both static and dynamic performances. Accordingly, a novel stability level index is put forward for stability comparison and mechanism analysis. Two inertia matching principles are then discovered. First, the inertia matching between the VSG and the SG is found to play a vital role in synchronization stability. Either too large or too small an inertia matching constant would lead to instability. There exists an optimal inertia matching value that guarantees the best stability performance. Second, the influence of the VSG penetration on the synchronization stability is demonstrated to be significantly dependent on the matching between the VSG inertia level and its voltage strength (i.e., output impedance). The coordination between the inertia and the virtual impedance helps to achieve robust stability. Regarding these two inertia matching principles, a control strategy based on inertia matching and virtual impedance adjustment of the VSG is proposed for synchronization stability enhancement and current suppression.

\bibliographystyle{ieeetr}
\bibliography{ref.bib}

\end{document}